\documentclass[prb,aps,amsmath,amssymb,nofootinbib,epsfig,showpacs,twocolumn,superscriptaddress]{revtex4-1}

\usepackage{amsmath,cleveref,amssymb,amsfonts,graphicx,multirow,color,bm,textcomp,mathtools}
\usepackage{paralist}
\usepackage{srcltx}
\usepackage[all,cmtip]{xy}
\usepackage{mathrsfs}
\usepackage{srcltx,soul,subfigure}
\usepackage{threeparttable,dsfont,bbm}

\crefname{equation}{Eq.}{Eqs.}
\crefname{figure}{Fig.}{Figs.}

\newcommand{\bra}[1]{\langle #1 |}
\newcommand{\ket}[1]{| #1 \rangle}

\newcommand{\bee}{\begin{eqnarray}}
\newcommand{\ee}{\end{eqnarray}}
\newcommand{\bma}{\begin{pmatrix}}
\newcommand{\ema}{\end{pmatrix}}
\newcommand{\balig}{\begin{align}}
\newcommand{\ealig}{\end{align}}

\newcommand{\ba}{\begin{align}}
\newcommand{\ea}{\end{align}}

\newcommand{\ignore}[1]{}

\usepackage{dcolumn}
\newcolumntype{C}[1]{>{\centering\let\newline\\\arraybackslash\hspace{0pt}}m{#1}}

\begin{document}

\title{Fractional Josephson Effect with and without Majorana Zero Modes}

\author{Ching-Kai Chiu}
\affiliation{
Condensed Matter Theory Center and Joint Quantum Institute and Station Q Maryland, Department of Physics, University of Maryland, College Park, MD 20742, USA}
\affiliation{Kavli Institute for Theoretical Sciences, University of Chinese Academy of Sciences, Beijing 100190, China}

\author{S. Das Sarma}
\affiliation{
Condensed Matter Theory Center and Joint Quantum Institute and Station Q Maryland, Department of Physics, University of Maryland, College Park, MD 20742, USA}


\begin{abstract} 

	It is known that the low-energy physics of the Josephson effect in the presence of Majorana zero modes exhibits a $4\pi$ periodicity as the Aharonov-Bohm flux varies in contrast to the $2\pi$ Josephson periodicity in usual superconducting junctions. We study this fractional Josephson effect in 1D topological superconductors in Majorana nanowire systems by focusing on the features of the phase-energy relations in a superconducting semiconductor nanowire with spin-orbital coupling by including different factors operational in experimental systems, such as short wire length, suppression of superconducting gap, and the presence of an Andreev bound state. We show that even in the absence of the Majorana zero modes, some non-topological physical effects can manifest a $4\pi$ periodicity of the phase-energy relation in the Josephson junction, thus providing an alternative physics for fractional Josephson effect with no underlying Majorana zero modes. Furthermore, we consider several scenarios of inhomogeneous chemical potential distributions in the superconducting nanowire leading to four Majorana bound states and construct the effective four Majorana model to correctly describe the low-energy theory of the Josephson effect. In this setup, multiple Majorana zero modes can also have the $4\pi$ fractional Josephson effect, although the underlying physics arises from Andreev bound states since two close by Majorana bound states effectively form Andreev bound states. Our work demonstrates that the mere observation of a fractional Josephson effect simulating $4\pi$ periodicity might not, by itself, be taken as the definitive evidence for topological superconductivity. This finding has important implications for the ongoing search for non-Abelian Majorana zero modes and efforts for developing topological qubits. 
	
\end{abstract}

\date{\rm\today}
\maketitle

\section{introduction}

	Topological phases of matter\cite{QiZhang2011,HassanKane2010,RevModPhys.88.035005} possess robust boundary states protected by topology against weak perturbations in bulk gapped systems and have attracted great attention in the condensed matter community. 
	In particular, in a topological superconductor (TSC) novel Majorana zero modes\cite{Kitaev2001,elliott_franz_review,Mourik_zero_bias,RokhinsonLiuFurdyna12,Deng_zero_bias,Churchill_zero_bias,Das_zero_bias,Finck_zero_bias,Albrecht:2016aa} (MZMs), which are their own antiparticles, have arisen from the encouraging experimental progress\cite{Mourik_zero_bias,RokhinsonLiuFurdyna12,Deng_zero_bias,Churchill_zero_bias,Das_zero_bias,Finck_zero_bias,Albrecht:2016aa} during the past six years. One  of  the particularly promising platforms \cite{FuKane_SC_STI,Franz_nanowire} to host and probe MZMs is a superconducting proximitized semiconductor nanowire with spin orbital coupling and adjustable Zeeman splitting\cite{Sau_semiconductor_heterostructures,Roman_SC_semi,Gil_Majorana_wire}. It is theoretically well-established that the combination of the spin orbital coupling and the Zeeman spin splitting converts an ordinary s-wave superconductor (SC) into an effective spinless p-wave TSC provided the Zeemn splitting is large enough (with the SC gap, protecting at the MZMs localized at the two ends of the nanowire and being proportional to the spin orbital coupling strength). The recent observation of the $2e^2/h$ quantized conductance peak~\cite{PhysRevB.63.144531,Zhang:2018aa} at zero bias voltage in this nanowire setup is a significant breakthrough providing support for the existence of MZMs. But, the zero bias tunnel conductance peak is only a necessary condition for MZMs, and cannot decisively establish their existence. In addition to the zero bias conductance peak, a TSC with MZMs should also manifest the so-called fractional Josephson effect, where the Josephson periodicity is $4\pi$ rather than $2\pi$\cite{Kwon2004,Kitaev2001,Roman_SC_semi}. The current work takes a deeper look at the fractional Josephson effect physics of MZM-carrying TSC, and shows that the $4\pi$ fractional Josephson effect is also merely a necessary, but not a sufficient, condition for establishing the MZM existence, and as such experimental claims for the manifestation of any fractional Josephson effect must also be treated with caution in this context. 
	
	
	Controlling MZMs to build fault-tolerant quantum gates is one of the leading directions to achieve quantum computation\cite{RMP_braiding}. 
	However, it has been established\cite{PhysRevA.73.042313,PhysRevB.73.245307} that although braiding MZMs offers the topological protection of the Majorana qubits, without additional non-topological gates this MZM braiding scheme cannot achieve universal quantum computation. Hence, for quantum computation, the topological protection has to be sacrificed to some extent by tuning the couplings of MZMs in order to manipulate the Majorana qubits and to make dense gate operations computationally. It has been proposed that adjusting the magnetic flux in the Josephson junction hosting MZMs can experimentally achieve the tuning of the MZM couplings~\cite{2016arXiv161005289K,PhysRevB.88.035121,Chiu_oneD_Majorana_move,1367-2630-14-3-035019} in these gate operations, and the Majorana Josephson junction is recognized as one of the basic building blocks to construct quantum gates. Therefore, understanding the physics of the Majorana Josephson junction is an important task toward quantum computation. 
	
	It was predicted by Kitaev~\cite{Kitaev2001} that the Majorana Josephson effect with a $4\pi$ periodicity as a function of the magnetic flux ($\Phi$) should emerge in an idealized model of a spinless $p$-wave topological superconductor carrying MZMs. Since the direct experimental realization of a spinless $p$-wave superconductor is challenging, alternative setups for the Majorana Josephson junctions have been theoretically proposed~\cite{Kwon2004,PhysRevB.92.134516,PhysRevB.93.184502,PhysRevB.94.115430,ZhangKane14,0953-8984-24-32-325701} and experimentally studied\cite{2017arXiv171208459L,WilliamsGoldhaber12,RokhinsonLiuFurdyna12,YamakageTanaka13,Kurter:2015aa} in the literature. Although in these experiments the $4\pi$ periodicity was observed, other non-topological factors, such as the suppression of the superconducting gap, can also manifest an effective $4\pi$ periodicity in the absence of MZMs\cite{PhysRevLett.54.2696,PhysRevLett.55.1610,PhysRevLett.56.386}; therefore, the observation of the $4\pi$ periodicity might be not decisive to conclude the existence of the topological superconductors hosting MZMs. In fact, the experimental situation is confusing here with the claimed observations of the $4\pi$ Josephson effect not reproduced typically and not fitting well generally with theoretical expectations.  
	
\begin{figure}[t!]
\begin{center}
\includegraphics[clip,width=0.78\columnwidth]{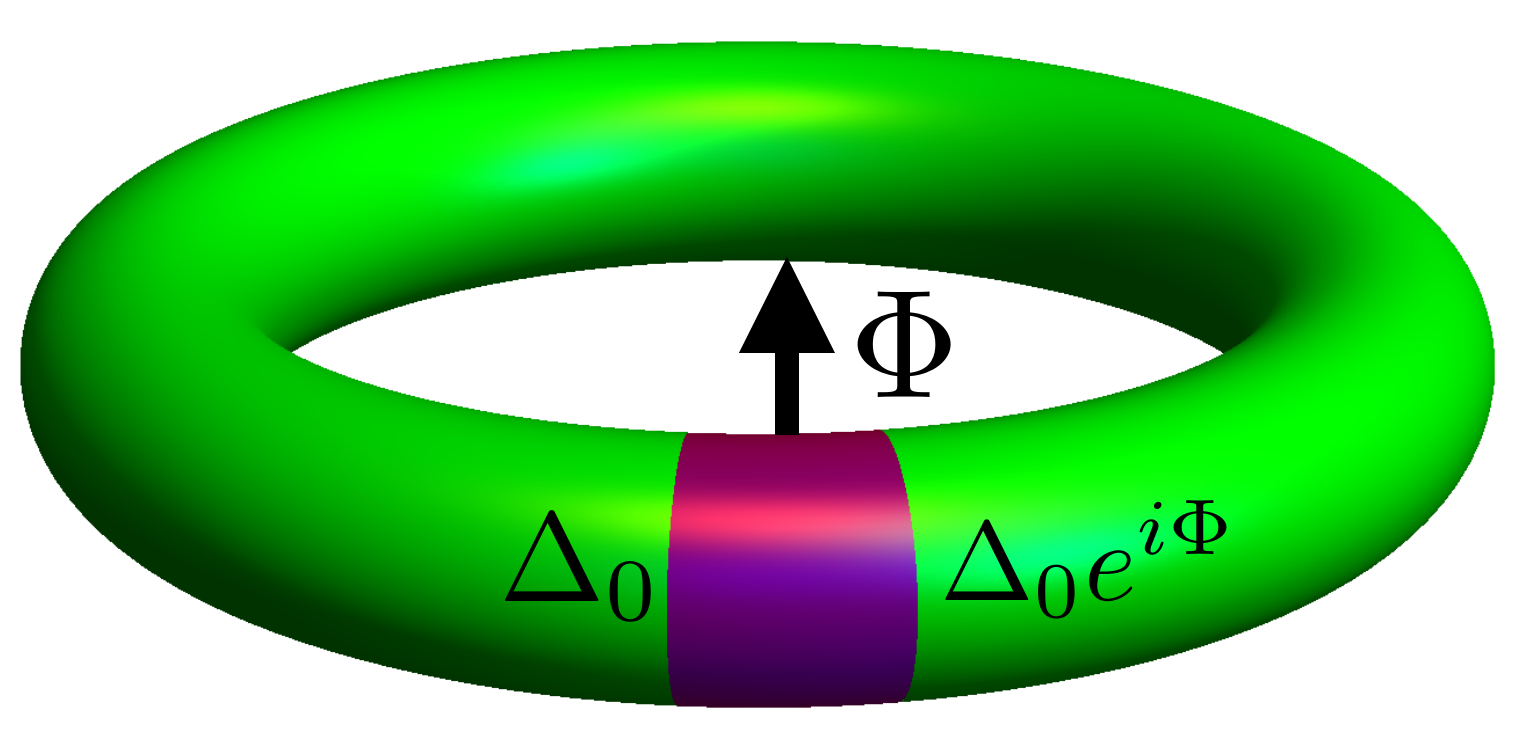}
  \caption{  The schematic of the Josephson junction considered in the theory. The green region represents the superconducting nanowire ($1\le j \le L$), while the purple color represents a thin junction so that the effective low energy physics can be described by a weak coupling between the two wire ends. The magnetic flux $\Phi$ through the middle of the ring leads to the phase difference between the two ends of the superconducting nanowire. We study the $\Phi-E$ relation in this system incorporating various physical effects in the presence or absence of MZMs. } 
  \label{schematic}
  \end{center}
\end{figure}

	In this manuscript, we theoretically examine non-trivial and trivial TSC Josephson junctions mediated by a superconducting proximitized semiconductor nanowire with and without MZMs in-depth. A detailed theoretical analysis of the TSC Josephson junctions associated with semiconductor nanowires is timely given the great current interest in this system as a platform for quantum computation and in the context of the recent observation of the MZM conductance quantization in the nanowires~\cite{2017arXiv171208459L}.
	As shown in Fig.~\ref{schematic} the Josephson junction setup considered in our work, the wire-junction system encloses a magnetic flux as usual, leading to the Aharonov-Bohm phase. The two ends of the superconducting nanowire with different Aharonov-Bohm phases \emph{weakly} couple as a Josephson junction due to the coherent tunneling of electrons. The difference between the Aharonov-Bohm phases in the wire ends, which stems from the magnetic flux ($\Phi$), leads to an oscillation of the quasiparticle and quasihole energy spectrum. The hybridized MZMs in the wire ends lead to the low energy phase-dependent spectrum $E \propto \cos (\Phi/2)$, exhibiting a $4\pi$ periodicity \cite{Kitaev2001,Kwon2004}. In reality, the experimental setup of the trivial and non-trivial TSCs might possess unavoidable physical effects affecting the periodicity in $\Phi$, such as short wire length, superconductor gap suppression, the presence of an Andreev bound state (ABS), and inhomogeneous potential distribution. 
	By separately including various physical mechanisms related to the $2\pi$ and $4\pi$ periodicities, we obtain theoretically the phase-energy ($\Phi-E$) relations to understand the physics of the Josephson effect with and without MZMs. Moreover, to circumvent non-essential complexity, we consider Landau-Zener tunneling\cite{Landau_tunneling,Zener_tunneling} to be absent by assuming that the system always adiabatically evolves at zero temperature with the conservation of fermion parity as the magnetic flux varies.

	The remainder of this paper is organized as follows. In sec.~\ref{continuous model}, we derive the low energy physics in the continuum model of the Josephson effect hosting two MZMs separately at the wire ends of the topological superconductor. We show that tunneling through the junction leads to $4\pi$ periodicity, and that the finite size effect of the short wire length manifests $\Phi$-independent energy splitting of these MZMs. Sec.~\ref{setup} provides the Josephson junction lattice model of the superconducting semiconductor nanowire with realistic physical parameters. In sec.~\ref{wire}, by using the lattice model, we calculate the $\Phi-E$ relation of superconducting semiconductor Josephson junctions in separate situations with distinct physical constraints: short wire length, superconducting gap suppression, and the presence of an ABS induced in a quantum dot. In sec.~\ref{long trivial}, we consider the realistic experimental setup\cite{2017arXiv171208459L} with a long conventional superconductor in the middle of wire and the topological superconductors are on the two sides of the trivial superconductor. We examine if the trivial cases leading to the $4\pi$ periodicity can be excluded by this setup of the long trivial superconductor.  Sec.~\ref{inh} is devoted to the $\Phi-E$ relation in the presence of inhomogeneous potentials, which produce two separate topological regions in the wire hosting multiple MZMs. Finally, in Sec.~\ref{conclusion} we summarize the various factors affecting the periodicity of the Josephson junction. We note that we have considered some of the more important physical mechanisms affecting the Josephson junction MZM physics in nanowires, and the there may vary well be other factors, not considered in our work, which could affect the TSC Josephson effect in nanowires.

\section{Continuum model for fractional Josephson Effect}\label{continuous model}



	Before considering specific models of Josephson junctions mediated by the superconducting-proximitized nanowire, we first use the continuum theory of a TSC to show that the $\Phi-E$ relationship is simply proportional to $\cos(\Phi/2)$ leading to a $4\pi$ periodicity. This relation stems from the hybridized MZMs in the TSC near the two wire ends. Furthermore, since the nanowire forming the junction is invariably of finite length leading to MZM wavefunction overlap from the two ends, the hybridization of the MZMs in the wire manifests energy splitting\cite{PhysRevLett.103.107001,PhysRevB.86.220506}.
	We show that this MZM overlap energy splitting is $\Phi$-independent.
	Our derivation scheme is in the following. First, by considering the superconducting nanowire in the open boundary condition, we find the MZM wavefunctions at the nanowire ends as domain walls. Then in the presence of the magnetic flux $\Phi$ we turn on the weak coupling between the two nanowire ends as a first order perturbation. This perturbation energy as the effective lowest energy is the key for the Majorana Josephson junction.  We further show that a MZM in one wire end is not an exact eigenstate at the other end so that the two Majorana  end modes hybridize strongly through the wire. We use the fundamental unit of magnetic flux, $\phi_0=h/2e$, as the unit of flux so that the flux $\Phi$ is the same as the phase difference between the two wire ends in our notation, and a phase function $\cos \Phi (\cos \Phi/2)$ automatically implies $2\pi(4\pi)$ periodicity in the Josephson effect.

	We start with the Bogoliubov-de-Gennes (BdG) Hamiltonian of the 1D superconducting semiconductor nanowire in momentum space~\cite{Sau_semiconductor_heterostructures,Roman_SC_semi,Gil_Majorana_wire}  
\bee
H_{\rm{BdG}}(k)=& \big[ 2t(1- \cos k ) - \mu \big] \tau_z \sigma_0 + \Delta_0 \tau_y \sigma_y \nonumber \\
&+ V_z \tau_z \sigma_z  + 2\alpha \sin k \tau_z \sigma_y, \label{k lattice}
\ee	
where Pauli matrices $\tau_\alpha$ and $\sigma_\beta$ represent the usual particle-hole and spin-$1/2$ degrees of freedom respectively, the superconducting order parameter $\Delta_0$ is a positive constant, $t$ is the strength of the nearest neighbor hopping, $\mu$ chemical potential, $V_z$ Zeeman splitting energy, $\alpha$ spin-orbit coupling; we further choose the lattice constant $a\equiv 1$. For large Zeeman splitting $V_z$, the superconducting nanowire is in the topological phase hosting MZMs at the wire ends. We are interested in the low energy theory near the Fermi level as $\mu\approx 2t (1 - \cos k)$. The BdG Hamiltonian can be further simplified in the continuum approximation 
\bee
H_{\rm{BdG}}(k)\approx 2\alpha k\tau_z \sigma_y  + \Delta_0\tau_y \sigma_y + V_z \tau_z \sigma_z.  \quad
\ee 
Since our focus is on the Majorana bound states near the nanowire boundaries ($x=0,\ L$, where $L$ is the nanowire length), for $0\leq x \leq L$ the phase of the superconducting order parameter $\Phi_x=\Phi x/L$ is position-dependent, where $\Phi$ indicates the magnetic flux through the wire. That is, $\Phi_{x}=0$ for $x= 0$ and $\Phi_x=\Phi$ for $x= L$. With this additional flux-induced phase, the order parameter is changed to $\Delta_0 c^\dagger_{x\uparrow} c^\dagger_{x\downarrow} \rightarrow \Delta_0 e^{i\Phi_x} c^\dagger_{x\uparrow} c^\dagger_{x\downarrow}$ and the momentum for particles and holes have different transformations $k\rightarrow k- eA, k+ eA$ respectively, where the magnetic potential $A=\Phi/2eL$ for $0\leq x \leq L$. The low energy Hamiltonian with the magnetic flux $\Phi$ can be written as 
\begin{align}
H_{\rm{BdG}}(x) =& 2\alpha ( \tau_z \sigma_y  \frac{\partial}{i\partial x} -eA\tau_0 \sigma_y)+\Delta_0 \cos \Phi_x \tau_y \sigma_y  \nonumber \\
&+ \Delta_0 \sin \Phi_x \tau_x \sigma_y  + V_z \tau_z \sigma_z ,
\end{align}
where we assume $\hbar\equiv 1$. The two wire ends are located at $x=0,\ L$ as domain walls (the wire boundary is equivalent to a domain wall), which follow $V_z$ is a constant ($V_{z0}$) and $ V_z  - \Delta_0 >0 $ for $0<x<L$ and $V_z  =0 $ elsewhere. That is, we introduce the trivial region ($x<0$ and $x>L$) sandwiching the topological region ($0\leq x \leq L$). The two important parameters are given by $A=0,\ \Phi_x=0$ for $x<0$ and $A=0, \Phi_x=\Phi$ for $x>L$. By solving the eigenvalue problem at zero energy, we have a normalizable Majorana wavefunction with zero energy localized near $x=0$ 
\bee
\ket{\phi_0(x)}=e^{-\frac{V_z-\Delta_0}{2\alpha}x}
\bma 
i e^{i\Phi_x/2}\\
-i e^{i\Phi_x/2}\\
-i e^{-i\Phi_x/2}\\
i e^{-i\Phi_x/2}\\
\ema.
\ee
In addition, another normalizable Majorana wavefunction with zero energy localized near $x=L$ is written as
\bee
\ket{\phi_L(x)}=e^{\frac{V_z-\Delta_0}{2\alpha}(x-L)}
\bma 
e^{i\Phi_x/2} \\
e^{i\Phi_x/2} \\
e^{-i\Phi_x/2} \\
e^{-i\Phi_x/2} \\
\ema.
\ee
We turn on the coupling between the two ends as the extension of spin orbital coupling
\bee
\Delta \hat{h}= i\delta ( C^\dagger_0 \tau_z \sigma_y C_L -C^\dagger_L \tau_z \sigma_y C_0 ).
\ee  
This coupling is weak enough to be a first order perturbation so that the low energy effective Hamiltonian can be written as the coupling sandwiched by the two MZMs 
\begin{align}
\Delta H =& \bma 
\bra{\phi_0} \Delta \hat{h} \ket{\phi_0} & \bra{\phi_0} \Delta \hat{h} \ket{\phi_L} \\
 \bra{\phi_L} \Delta \hat{h} \ket{\phi_0} & \bra{\phi_L} \Delta \hat{h} \ket{\phi_L}
\ema \nonumber \\
\propto & 4\delta \cos (\Phi/2)
\bma
0 & -i \\
i & 0 \\
\ema
\end{align}
The energy spectrum $\Delta E \propto \pm \cos (\Phi/2) $ seems to have a $2\pi$ periodicity. 
However, as the magnetic flux $\Phi$ varies from $0$ to $4\pi$, by following the adiabatic evolution of a hybridized Majorana state with fixed parity (say $\cos( \Phi/2)$) the system exhibits a $4\pi$ periodicity, which is the key idea of this manuscript. 


	If the wire is too short, the hybridization of the Majorana bound states leads to large $\Phi$-independent energy splitting. To show this splitting, we start with the wavefunction $\ket{\phi_L}$, which is the exact solution of $H_{\rm{BdG}}(x)$ for $x>0$. 
	The tail of $\ket{\phi_L}$ at $x<0$ cannot be a part of the eigenstate of $H_{\rm{BdG}}(x)$, since the rapid change of the Zeeman splitting is located at $x=0$ as a domain wall. This non-vanishing part for $x<0$ is given by 
\bee	
H_{\rm{BdG}}\ket{\phi_L(x)}= -V_{z0} \tau_z \sigma_z \ket{\phi_L(x)}.
\ee
This leads to the effective coupling between the two MZMs in the wire
\begin{align} 
\bra{\phi_0} H_{\rm{BdG}} \ket{\phi_L}&=\int dx \bra{\phi_0(x)} H_{\rm{BdG}} \ket{\phi_L(x)} \\
&=4iV_ze^{-\frac{V_z-\Delta_0}{2\alpha}L}\int_{-\infty}^0 e^{\frac{V_z}{2\alpha}x}dx.
\end{align} 
The Majorana hybridization is given by  
\begin{align}
\Delta H' =& \bma 
\bra{\phi_0} H_{\rm{BdG}} \ket{\phi_0} & \bra{\phi_0} H_{\rm{BdG}} \ket{\phi_L} \\
 \bra{\phi_L} H_{\rm{BdG}} \ket{\phi_0} & \bra{\phi_L} H_{\rm{BdG}} \ket{\phi_L}
\ema \nonumber \\
\propto & -8\alpha e^{-\frac{V_z-\Delta_0}{2\alpha}L}
\bma
0 & -i \\
i & 0 \\
\ema.
\end{align}
Hence, this energy splitting is $\Phi$-independent due to the finite size effect. 
Including the Majorana hybridization through the junction, the total energy splitting is written as
\bee 
\Delta E_\pm=\pm d \cos (\Phi/2) \pm D \label{theory splitting}.
\ee
Hence, the $\Phi-E$ relation exhibits a $4\pi$ periodicity independent of MZM overlap. This low energy theory can be further written in the Majorana basis
\bee
H_{\rm{2M}}=i(d\cos(\Phi/2)+D)\gamma_l\gamma_r \label{M2},
\ee
where $\gamma_l,\ \gamma_r$ are Majorana operators at the two wire ends. 
We will show that in sec.~\ref{wire} the energy splitting (\ref{M2}) is the key simply describing the phase-energy relation in most cases of the \emph{fractional} Josephson junction. We note that the Josephson periodicity here is $4\pi$ even when the two MZMs overlap strongly in a short nanowire leading to large $\Phi$-independent energy splitting ($D$), but such a short wire is unsuitable for topological gate operations since such overlapping MZMs are not non-Abelian objects.

\section{simulation setup}\label{setup}

The simulation setup to study the phase-energy relation of the Josephson junction is based on the nanowire model described by the approximate experimental parameters.
We start with the lattice Hamiltonian (obtained by discretizing Eq.~\ref{k lattice}) in the ring geometry as shown in Fig.~\ref{schematic}
\begin{small}
\begin{align}
\hat{H}_{\rm{BdG}}=&\sum_{1\leq j \leq L } \Big \{  C^\dagger_j  \big [ (2t-\mu)\tau_z \sigma_0 +\Delta_0\tau_y\sigma_y + V_z \tau_z \sigma_z \big ]C_j, \nonumber \\
&+\big [ C_{j+1}^\dagger  (-t \tau_z \tau_0 + i \alpha \tau_z \sigma_y )C_j + wC^\dagger_L \tau_z \sigma_0  C_1 + \rm{h.c.} \big ]  \Big \}, \label{lattice Hamiltonian}
\end{align}
\end{small}
where the Nambu spinor is given by $C_j = (c_{\uparrow j}, c_{\downarrow j}, c^\dagger_{\uparrow j }, c^\dagger_{\downarrow j})^T$.
For the Josephson junction, the wire ends are commonly separated by a thin insulator or metal. To simplify the problem, we assume these ends directly couple through the weak tunneling with strength $w$.
For $w=0$, the nanowire in the open boundary condition may host MZMs at the two ends if the system parameters are in the appropriate topological regime (i.e. $V_z$ large enough compared with $\mu$ and $\Delta$). The non-zero tunnel coupling $w$ hybridizes these two Majorana end modes through the junction lifting their energies away from zero because of Majorana splitting. To appropriately describe the current experimental setup, the values of the physical parameters\cite{PhysRevB.96.075161} (unless specified otherwise) are taken to be lattice constant $a=10$nm, hopping strength $t=25$meV, spin-orbital coupling $\alpha=2.5$meV, superconducting gap $\Delta=0.9$meV, chemical potential $\mu=4$meV, tunneling of the two wire ends $w=0.1$meV. 
By comparing with the low-energy effective Hamiltonian, 
\begin{align} 
H_{\rm{low E}}(x)= - \frac{\hbar^2}{2m^*}{\partial_x^2} \tau_z \sigma_0 + i\alpha_l  \partial_x +\Delta_0 \tau_y \sigma_y + V_z \tau_z \sigma_z, 
\end{align}
the values of the corresponding parameters are written as the spin-orbital coupling $\alpha_l=0.5$eV$\cdot \AA$ and the effective mass $m_{\rm{eff}}=0.016m_e$, where $m_e$ is the electron rest mass. Although we use these parameters for our numerical simulations, obviously the qualitative features of our results do not depend on any specific parameter choice.

	Now we introduce the magnetic flux $\Phi$ in the unit of flux quantum $h/2e$ going through the middle of the nanowire as shown in Fig.~\ref{schematic}. The non-superconducting junction keeps the phase difference between the two ends being $\Phi$ and then the superconducting order parameter has an additional position-dependent phase    
\begin{align}
\Delta_0 & C^\dagger_j \tau_y \sigma_y C_j \rightarrow \Delta_0 C^\dagger_j\big [ \cos (2j \phi)\tau_y \sigma_y + \sin (2j \phi)\tau_x \sigma_y \big ] C_j,
\end{align}
where $\phi=\Phi/2L$ indicates the phase difference between the two nearest neighbor sites. Furthermore, the nanowire Hamilton is modified by the Peierls substitution in the presence of the applied flux.
\begin{small}
\begin{align}
C^\dagger_{j+1}& (-t \tau_z \sigma_0 + i \alpha \tau_z \sigma_y )C_j \nonumber \\
&\rightarrow C^\dagger_{j+1}\big [ (\tau_z+\tau_0)e^{-i\phi}+ (\tau_z -\tau_0)e^{i\phi}\big ] \frac{-t\sigma_0 + i\alpha \sigma_y}{2} C_j,
\end{align}	
\end{small}
 With this construction, we numerically solve the eigenvalues of the lattice Hamiltonian in the presence of the applied flux to obtain the energy spectrum and the $\Phi-E$ relation for the nanowire Josephson junction.

\section{Energy Spectrum and $\Phi-E$ relation} \label{wire}

	We start with the Josephson junction of the original superconducting semiconductor nanowire lattice model and plot the $E-\Phi$ relation to show the Majorana Josephson effect exhibiting a $4\pi$ periodicity in $\Phi$. The conditions of lattice model are further extended to include short length wire, superconducting gap suppression, and the presence of an Andreev bound state to discuss the other mechanisms leading to a $4\pi$ periodicity in the absence of MZMs. Below we present these results sequentially including each physical mechanism individually. 


\subsection{Long superconducting semiconductor nanowire} \label{Topo_SC_sec}
	
\begin{figure}[t!]
\begin{center}
\includegraphics[clip,width=0.98\columnwidth]{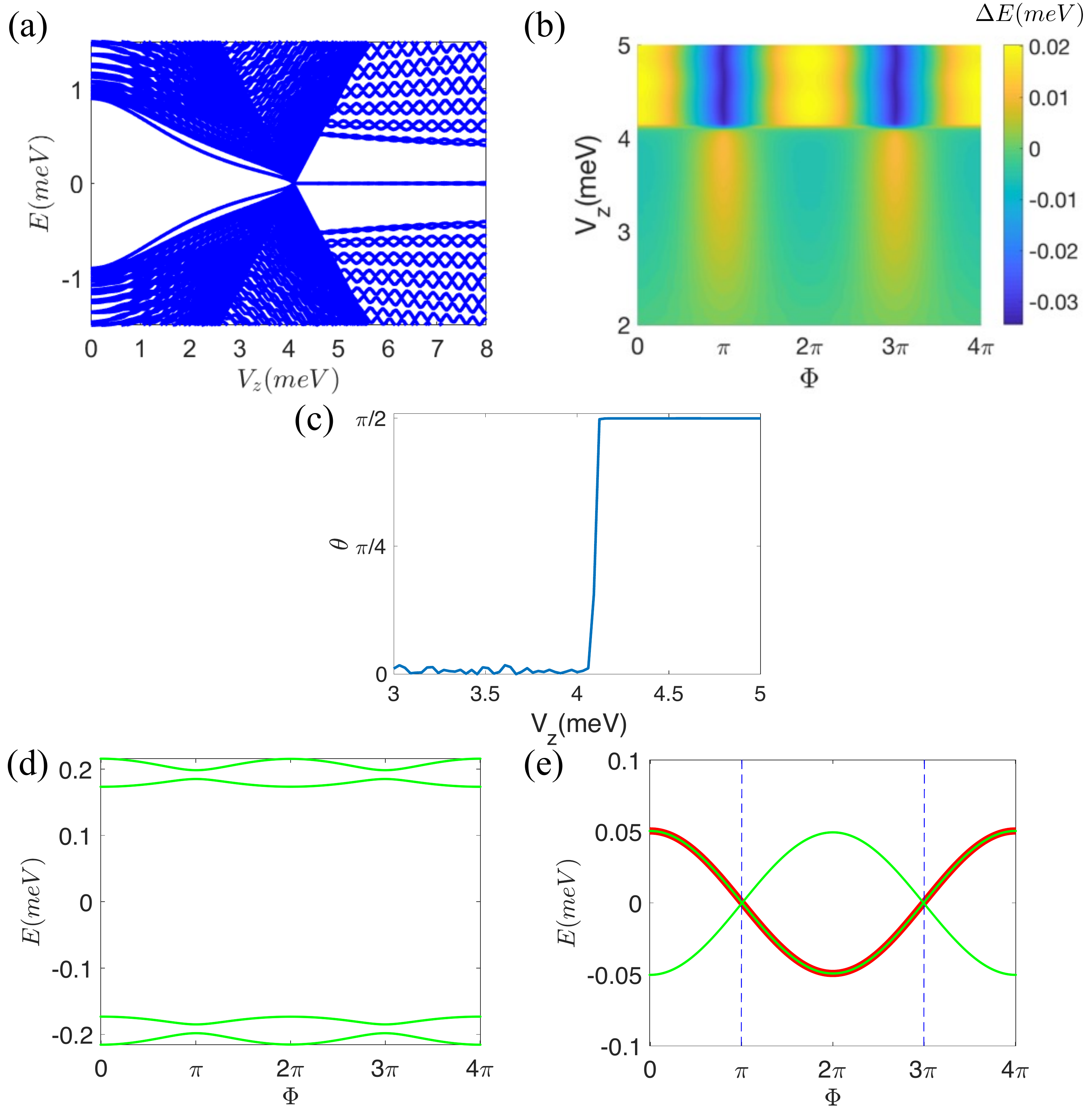}
  \caption{  (a) the energy spectrum of the lattice Hamiltonian in the open boundary condition showing that beyond the TQPT point ($V_z=\sqrt{\mu^2+\Delta^2_0}=4.1$meV), two MZMs appear on the two wire ends separately.  (b) The energy deviation as a function of $V_z$ and $\Phi$ is defined by $\Delta E (\Phi)\equiv E_1(\Phi)-\langle E_1 (\Phi) \rangle$, where $E_1$ is the lowest positive energy and $\langle E_1 (\Phi) \rangle$ is the average of $E_1$ from $\Phi=0$ to $4\pi$. The patterns sharply change at the TQPT point. The dark blue lines at $\Phi=\pi, 3\pi$ in the \emph{topological} region indicate the energy level crossings at zero energy. These crossings lead to the $4\pi$ periodicity in $\Phi$. (c) the angle $\theta=\arctan \rho_{4\pi}/\rho_{2\pi}$ shows the ratio between $2\pi$ and $4\pi$ periodicities obtained by the Fourier transformation of the lowest positive energy $E_1(\Phi)$, where  $\rho_{4\pi}$ and $\rho_{2\pi}$ are the strengths of $2\pi$ and $4\pi$ periodicities. The sharp jump at the TQPT indicates the transition from a $2\pi$ periodicity to a $4\pi$ periodicity. (d) the $\Phi-E$ relation at $V_z=3$meV in the trivial region shows a $2\pi$ periodicity. (e) the $\Phi-E$ relation (green) at $V_z=6$meV in the trivial region shows $4\pi$ periodicity and is consistent with the energy splitting $d\cos(\Phi/2)$ in the theory (red, $d$ is obtained by fitting). 
} 
  \label{Topo_SC}
  \end{center}
\end{figure}

	
	We consider the length of the nanowire to be long enough ($L=400$ lattice units $(4\mu\rm{m})$) so that the Majorana hybridization orginating from the finite size effect is strongly suppressed. In the absence of Zeeman splitting $(V_z=0)$, the nanowire in the trivial region does not possess MZMs at the two ends. As the Zeeman splitting increases, the system passes through the topological quantum phase transition (TQPT). For this specific model, the TQPT point is located at $V_z\approx \sqrt{\Delta^2+\mu_2^2}=4.1$meV.  After the TQPT, the MZMs with zero energy on the wire ends appear as shown in Fig.~\ref{Topo_SC}(a). The presence of these boundary MZMs indicates that the system is now a TSC with the bulk gap protecting the zero-energy (i.e. mid-gap) MZMs. 
	
	As the magnetic flux $\Phi$ goes through the middle of the ring, the two wire ends with the additional phases weakly couple. The energy deviation of the lowest positive energy (see the definition in the caption of Fig.~\ref{Topo_SC})  as a function of $V_z$ and $\Phi$ shows clear $2\pi$-periodicities in both the trivial and topological regions and has changes at the TQPT point  as shown in Fig.~\ref{Topo_SC}(b). In the trivial region, the lowest energy level never reaches zero as the magnetic flux varies from $0$ to $4\pi$ or touch the second energy level (Fig.~\ref{Topo_SC}(d)). In this regard, the BCS ground state adiabatically evolves back to the original state after  $2\pi$. However, in the topological region, at $\Phi=\pi, 3\pi$ the Majorana modes at the two ends completely decouple with the exact zero energy. Hence, after the lowest positive energy quasiparticle, which starts at $\Phi=0$, passes through zero energy at $\Phi=\pi$ as shown in Fig.~\ref{Topo_SC}(e), it possesses negative energy at $\Phi=2\pi$ and then evolves back to the original quasiparticle at $\Phi=4\pi$. Furthermore, the energy dispersion can be described by $\Delta E_{\pm}=\pm d \cos (\Phi/2)$ in Eq.~\ref{theory splitting} with $D=0$ due to the suppression of the finite size effect in the long wire. 
	
	Thus, in the topological region the BCS ground state with fixed fermion parity has a $4\pi$ periodicity. We further Fourier analyze the lowest energy with the fixed parity as shown in Fig.~\ref{Topo_SC}(c). In the trivial regime, the $2\pi$-periodicity dominates with a small mixing of the $4\pi$ oscillation, whereas only $4\pi$ periodicity appears in the topological region. The $2\pi$ and $4\pi$ periodicities can clearly distinguish the trivial and topological regions in this \emph{perfect} scenario of a long nanowire, as is already known. We show these results for our specific situations so that our findings below including realistic physical effects not included in the idealized model can be distinguished from this  perfect scenario.

\begin{figure}[t]
\begin{center}
\includegraphics[clip,width=0.98\columnwidth]{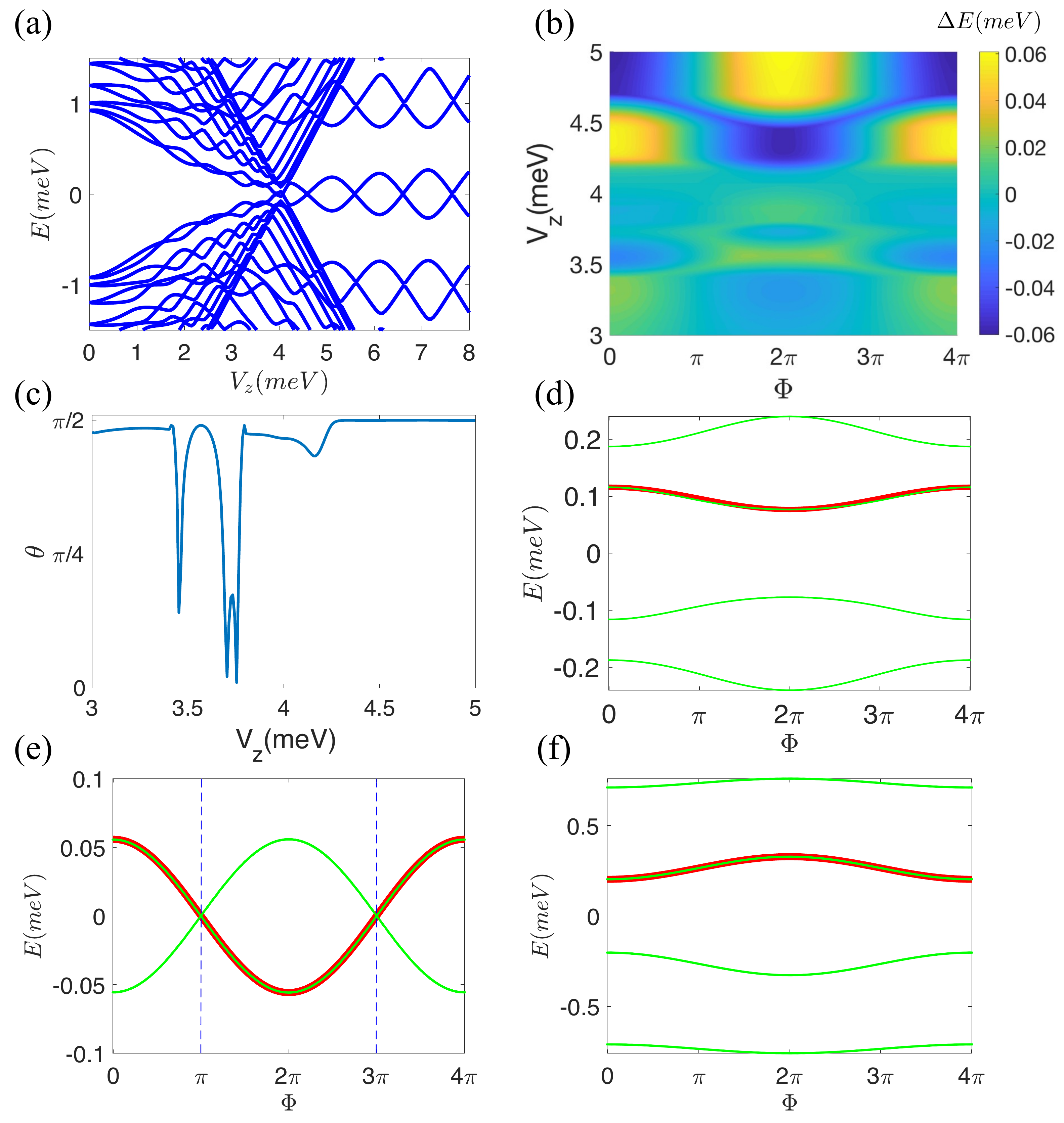}
  \caption{ (a) the energy spectrum for the short wire ($L=100$) shows that the Majorana bound states strongly hybridize with oscillations in the topological region. (b) the energy deviation as a function of $V_z$ and $\Phi$ shows that the $4\pi$ periodicity in $\Phi$ mostly dominates in the trivial and topological regions. (c) the ratio ($\tan \theta$) of the $4\pi$ and $2\pi$ periodicities shows in the trivial region the mixture of the $2\pi$ and $4\pi$ periodicities and as $V_z>4.26$meV the $2\pi$ periodicity  vanishes in the trivial regime. (d) the $\Phi-E$ relation (green) at $V_z=3.3$meV in the trivial region can be described by the effective two-Majorana theory (\ref{theory splitting}) with strong $\Phi-$indepedent Majorana hybridization. (e) $V_z=6.6$meV corresponding to zero energy leads to the lowest energy proportional to $\cos (\Phi/2)$.  (f) for $V_z=7.1$meV, Majorana end modes strongly hybridize through the wire so that the lowest energy ($\pm E_1$) shifts away from zero with $\pm\cos (\Phi/2)$ oscillation. The red lines in panels (d,e,f) presenting the Majorana effective theory ($D,\ d$ in Eq.~\ref{theory splitting} are chosen by fitting) are consistent with the simulation (green). } 
  \label{short_length}
  \end{center}
\end{figure}

\subsection{Short wires}
	
	We consider the experimental setup away from the perfect scenario above. For a short wire, the Majorana modes at the wire ends can easily hybridize away from zero energy due to the wavefunction overlap through the wire. In the topological region, the hybridization energy oscillates as the Zeeman splitting ($V_z$) is increased\cite{PhysRevB.86.220506,chiu_blockade,PhysRevLett.103.107001,PhysRevB.85.165124,PhysRevB.86.220506}. As shown in Fig.~\ref{short_length}(a), the wire spectrum in the open boundary condition shows the oscillation amplitude gradually is increased as $V_z$ increasing beyond the TQPT point. This arises from an effective increase of the superconducting coherence length because increasing $V_z$ reduces the SC gap energy. 
	
	In the entire system with the additional magnetic flux, the lowest positive energy fluctuation (Fig.~\ref{short_length}(b)) and the corresponding Fourier analysis (Fig.~\ref{short_length}(c)) manifest clear $4\pi$ periodicity in the topological region. In the trivial region, the $2\pi$ periodicity of the energy oscillation is seen mixed with the $4\pi$ periodicity. 
	Much of the trivial regime is in fact dominated by $4\pi$ oscillation. 
	In fig.~\ref{short_length}(d) at $V_z=3.3$meV in trivial region, even in the absence of the MZMs the lowest energy can be faithfully described by Eq.~\ref{theory splitting}, which is the model based on the hybridized MZMs. The clear emergence of the $4\pi$ periodicity in the trivial short wire is not surprising, 
since in the literature\cite{PhysRevB.33.5114} it has been shown that for a short conventional superconductor a $4\pi$ periodicity may be observed. Therefore, the observation of the $4\pi$ periodicity is not a conclusive evidence for topological superconductivity. It may very well be that one is dealing with a rather short trivial wire with no MZMs whose length is smaller than the SC coherence length as shown in Fig.~\ref{short_length}.

	In the topological region, when the $\Phi-$independent Majorana hybridization vanishes at few specific $V_z$ values, the oscillation of the lowest energy as a function of $\Phi$ (Fig.~\ref{short_length}(e)) is identical to the corresponding situation in the long wire limit (Fig.~\ref{Topo_SC}(e)). The zero energy modes appear at $\Phi=\pi,\ 3\pi$. On the other hand, due to the short length of the wire, the Majorana hybridization significantly affects the energy spectrum. With the hybridization energy $D$ as a $\Phi$-independent constant, the lowest energy in the simulation is in agreement with Eq.~\ref{theory splitting} as shown in Fig.~\ref{short_length}(f). That is, even when the MZMs are strongly hybridized, the junction still exhibits a $4\pi$ periodicity in the short wire. Although this finding seems unexpected, it is related to the coupling of the hybridized MZMs through the junction being proportional to $\pm \cos (\Phi/2)$. 

\begin{figure}[t!]
\begin{center}
\includegraphics[clip,width=0.98\columnwidth]{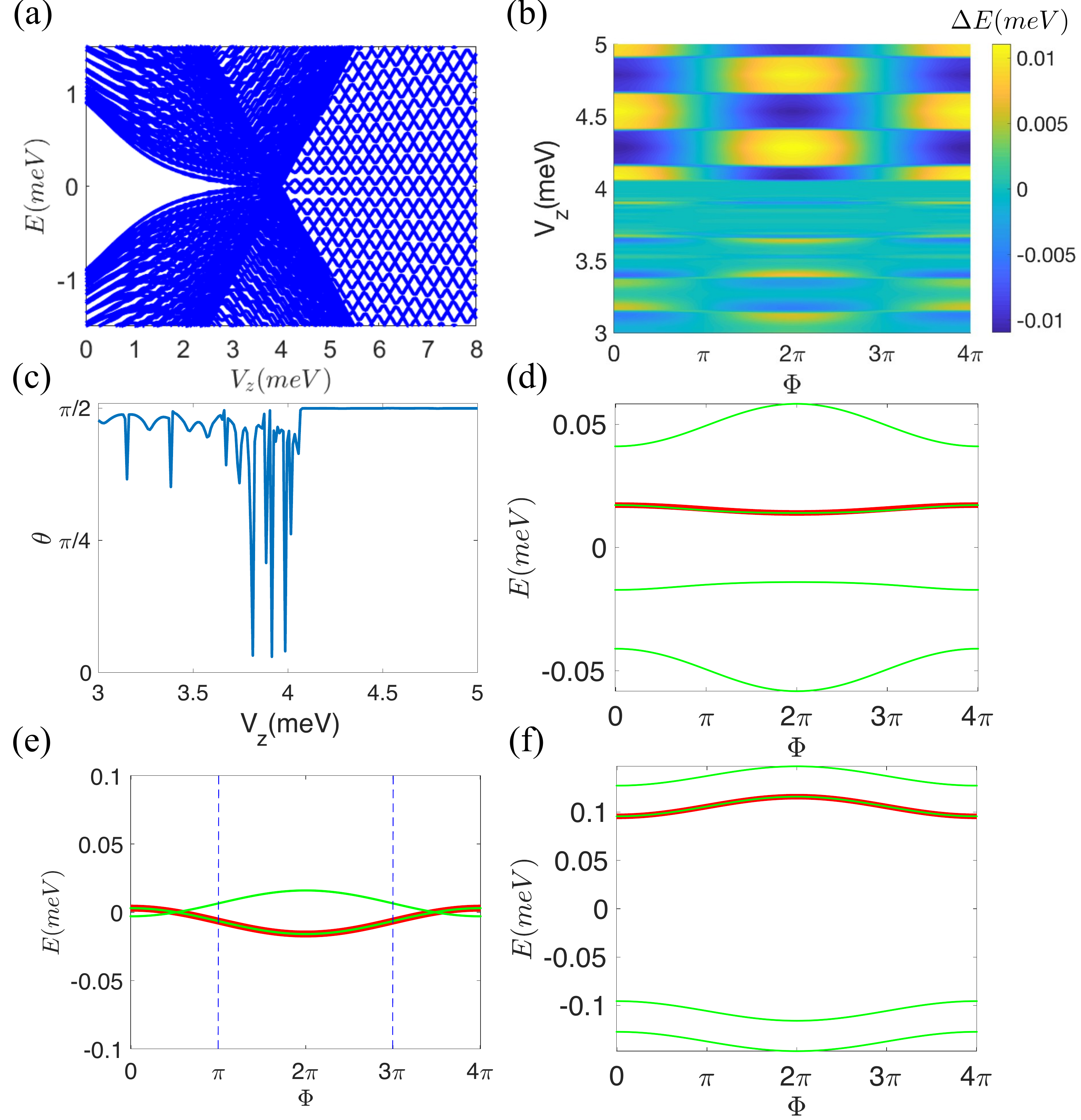}
  \caption{ (a) the energy spectrum in the presence of the suppression of the superconducting gap (\ref{SC gap}) shows the bulk gap closing at $V_z>4.1$meV, which is the TQPT point for the two previous cases. Due to the weak superconductivity, after the bulk gap closes, the lowest energy states are not localized MZMs. (b) the energy deviation shows the $4\pi$ periodicity domination in the trivial and topological regions. (c) the ratio of the $4\pi$ and $2\pi$ periodicities shows the $4\pi$ periodicity only after the bulk gap closing and the mixture of $2\pi$ and $4\pi$ periodicities before the bulk gap closing. Due to the small superconductor gap, $4\pi$ periodicity still dominates in the trivial region. 
  (d,e,f) the $\Phi-E$ relations (green) at $V_z=3.3,\ 4.4,\ 7.35$meV respectively can be described by the effective two-Majorana model (red, $D,\ d$ in Eq.\ref{theory splitting} are chosen by fitting). 
  } 
  \label{delta_decay}
  \end{center}
\end{figure}

\subsection{Superconducting gap suppression}

	Since the presence of the magnetic field might suppress the superconducting gap (e.g.~orbital effect) in the nanowire, we consider a model of the superconducting order parameter obeying exponential decay in the Zeeman splitting (the precise form of the decay function does not significantly affect the $\Phi-E$ relation.)
\bee	
\Delta=\Delta_0 e^{-V_z/\lambda}.	 \label{SC gap}
\ee
We intentionally choose $\lambda=2$meV so that the superconducting gap almost vanishes $V_z>4.1$meV, since $V_z=4.1$meV was the TQPT in the previous gapped superconducting systems. 
The spectrum of the nanowire in the open boundary condition (Fig.~\ref{delta_decay}(a)) shows that once that the bulk energy gap closes near the previous TQPT point ($V_z=4.1$meV), the bulk gap does not reopen as $V_z$ is increased ($\Delta \rightarrow 0$). Furthermore, the Majorana bound states, which hybridize, become a quasiparticle or a quasihole state with energy oscillation. For a small gap, in both the trivial and topological regions the system can be treated as an effective normal metal (particularly, at the finite experimental termperatures). It is not therefore surprising that as shown in Fig.~\ref{delta_decay}(b,c) the $4\pi$ periodicity behavior stems from the normal metal\cite{PhysRevLett.54.2696,PhysRevLett.55.1610,PhysRevLett.56.386}. It is worth by noting Fig.~\ref{delta_decay}(c) showing that the $2\pi$ periodicity completely vanishes after the TQPT point in this suppressed gap ``topological" regime. 

	In the trivial region before the bulk gap closing, the $\Phi-E$ relation (Fig.~\ref{delta_decay}(d)) can still be described by Eq.~\ref{theory splitting}. Similarly,  the $4\pi$ oscillation and the constant hybridization energy in Eq.~\ref{theory splitting} capture the $\Phi-E$  relation in the topological regions as shown in Fig.~\ref{delta_decay}(e,f).  As the magnetic field strongly suppresses the gap converting the system to an effective normal metal, the $4\pi$ periodicity of the junction is expected to occur since this is the ordinary Aharonov-Bohm oscillation in an ordinary metal. 



\begin{figure}[t!]
\begin{center}
\includegraphics[clip,width=0.98\columnwidth]{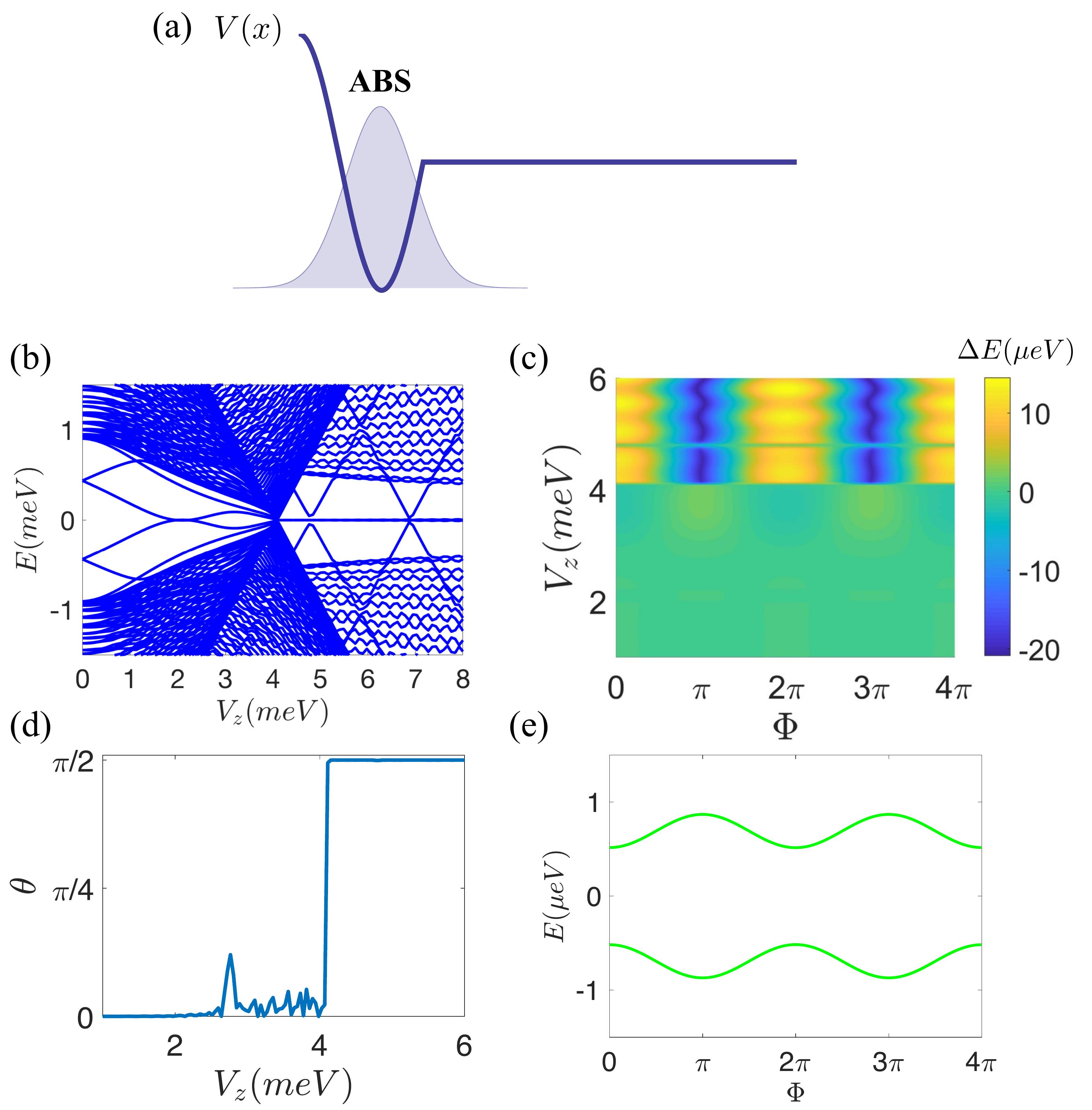}
  \caption{ (a) pictorially illustrates the additional potential dip at the wire end as the quantum dot leading to an ABS in the system. In the absence of the superconductivity, the quantum dot can host a low energy ABS. (b) the energy spectrum ($L=400$) with the quantum dot indicates that the low energy ABS appears for $V_z=1.8 \sim 2.4$meV.  (c) the energy deviation $\Delta E$ shows $2\pi$ periodicity in the trivial region although it might not be clear as $V_z$ moves away from the TQPT point since the region is all green. (d) the ratio ($\tan \theta$) of the $4\pi$ and $2\pi$ periodicities indicates that the $2\pi$ periodicity still dominates in the presence of the low energy ABS in the trivial region.  (e) the $\Phi-E$ relation at $V_z=2.2$meV with the low energy ABS exhibits a $2\pi$ periodicity. } 
  \label{localized_ABS}
  \end{center}
\end{figure}

\subsection{Quantum dot hosting an ABS near the wire end}

It is now well-known\cite{chiu_blockade,PhysRevB.96.075161} that many putative properties of MZMs in nanowires could be artificially simulated by accidental ABSs in nanowires which happen to be close to midgap in energy.  This is true for the zero bias conductance peak\cite{PhysRevB.96.075161} as well as the apparent MZM oscillations as a function of $V_z$ or $L$\cite{chiu_blockade}.  Each trivial ABS may be thought of as two spatially closely located MZMs in a varying background of chemical potential (caused, for example, by disorder or a quantum dot in the system), and if the experimental probe (e.g. the tunneling lead) couples strongly to only one of these MZMs, then the system response may mimic that of just an isolated MZM.  We now investigate the Josephson effect in our system by assuming the existence of a quantum dot induced ABS.
	It is likely that the presence of the ABS alters the $\Phi-E$ relation in the trivial phase. This motivates us to include a low energy ABS in the Josephson junction setup to see the features of the $\Phi-E$ relation by adding a quantum dot in one wire end. As illustrated in Fig.~\ref{localized_ABS}(a), an additional potential well, induced by a quantum dot (or other extrinsic mechanism), is included in the lattice Hamiltonian (\ref{lattice Hamiltonian}) 
\bee
\hat{H}_{\rm{well}}=\sum_{1\leq j \leq D} C_j^\dagger \cos (3\pi j/2L_D) \tau_z \sigma_0 C_j.
\ee	
For the simulation, the length of the quantum dot $L_D=30$ is used and the superconducting order parameter $\Delta$ is removed in this region (i.e.~the dot is assumed to be normal although this is not expected to be important for our results). The energy spectrum as a function of $V_z$ (Fig.~\ref{localized_ABS}(b)) shows the zero energy sticking of the ABS at the quantum dot as $V_z$ varies from $1.8$meV to $2.4$meV. Since the TQPT point is located at $V_z=4.1$meV, the low energy ABS is in the trivial regime and does not coexist with the MZMs. As shown in Fig.~\ref{localized_ABS}(c,d), even in the presence of the ABS, the $2\pi$ periodicity of the $\Phi -E$ relation still dominates in the topologically trivial region. By comparing with the trivial regime without the ABS (Fig.~\ref{Topo_SC}(c)), a small portion of $4\pi$ periodicity arises in the presence of the ABS in additional to the dominant $2\pi$ trivial oscillations. Nevertheless, the TQPT is still the main transition point between $2\pi$ and $4\pi$ periodicities.


\section{Inhomogeneous potentials} \label{inh}


	In the practical experimental setup, the homogeneous background potential as used so far in our simulations cannot be perfectly under control so that the inhomogeneous potential might lead to several disconnected topological regions in the nanowire with different adjacent spatial regimes separating into effective trivial and topological regimes with multiple nearby MZMs according to their local chemical potentials and superconducting gap values. Each topological region can host localized Majorana modes at its two ends, and in principle, depending on the details of the spatial inhomogeneity, there could be many MZMs located in the nanowire, not just two at the two physical boundaries at the wire ends. 
	In the following, to be specific we consider several inhomogeneous potential distributions, that have \emph{two} disconnected topological regions hosting four Majorana modes (i.e.~two MZMs in each spatial topological region), and then further study the $\Phi-E$ relations of the Josephson effect in the presence of these four MZMs. 
	

\subsection{Step function potential}

\begin{figure}[t!]
\begin{center}
\includegraphics[clip,width=0.98\columnwidth]{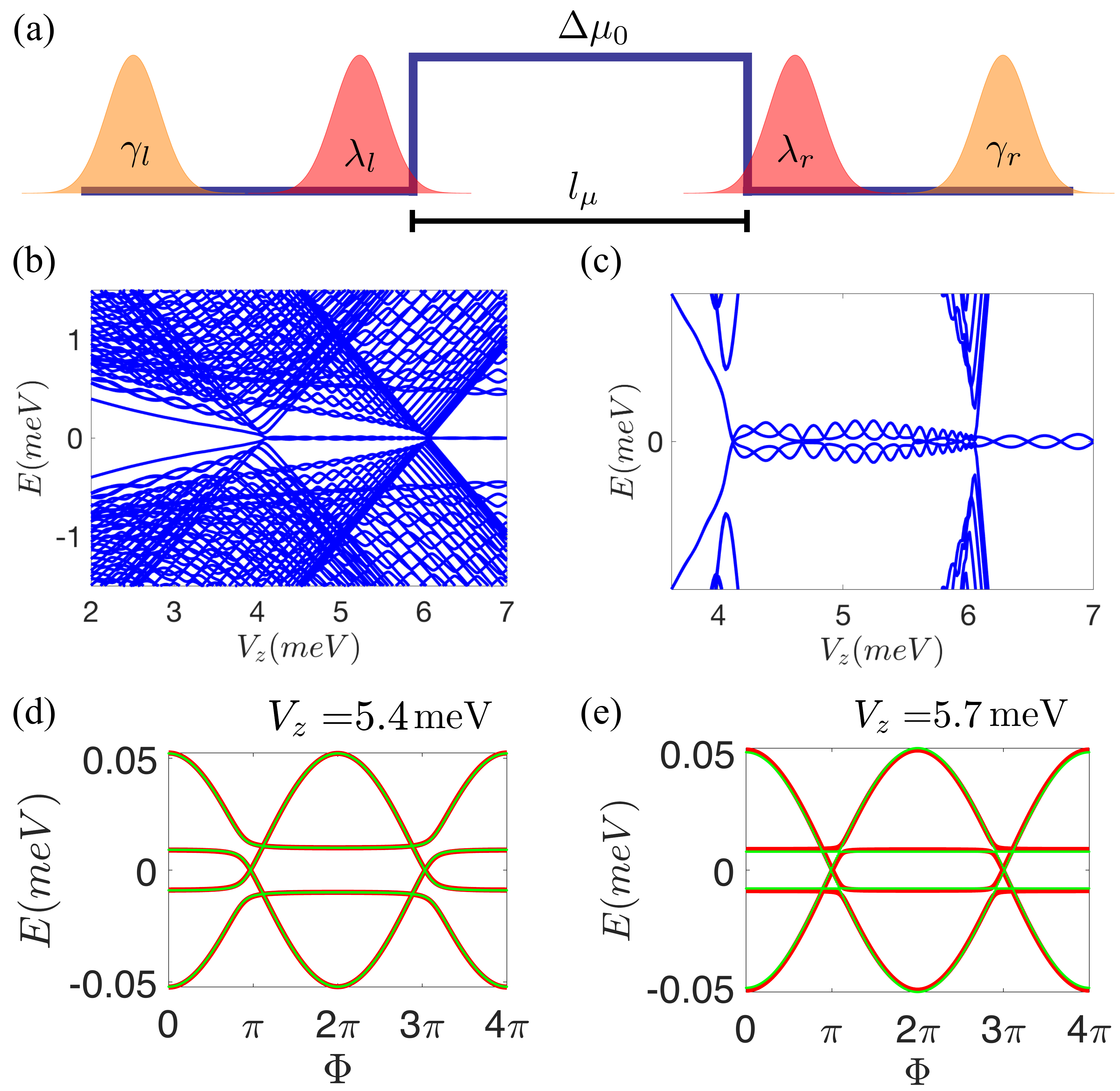}
  \caption{ (a) the potential distribution shows that an additional constant chemical potential $\Delta \mu_0$ is added in the middle of the wire with $l_\mu=200$. (b) the energy spectrum indicates that there are two TQPT points ($V_{z1}=4.1$meV and $V_{z2}=6.07$meV). (c) the spectrum in the small energy region indicates the presence of four Majorana modes ($\gamma_l,\ \lambda_l,\ \lambda_r,\ \gamma_r$) with energies close to zero for $V_{z1}<V_z<V_{z2}$. Their locations are illustrated in panel (a). As $V_z>V_{z2}$, only two MZMs ($\gamma_l, \gamma_r$) appear at the two wire ends respectively. (e,d) the $\Phi-E$ relations of the lattice model (green) for the four lowest energy levels are in agreement with the effective four-Majorana model (red, $d,\ f,\ g$ in Eq.~\ref{fourM_E} are chosen by fitting). The $4\pi$ periodicity domination persists in the entire four Majorana region. 
 } 
  \label{Step_function_lx200}
  \end{center}
\end{figure}

	We first consider the simplest inhomogeneous distribution of the chemical potential by adding just one constant potential in the middle region of the wire. For the simulation model, the constant potential well in the middle wire 
\bee
\hat{H}_{\rm{step}}=-\sum_{(L-l_\mu)/2 < j \leq (L+l_\mu)/2  } C_j^\dagger \Delta \mu_0  \tau_z \sigma_0 C_j, 
\ee	
is added in the lattice Hamiltonian (\ref{lattice Hamiltonian}). The reason for adding the negative potential (positive chemical potential) is that the two sides of the wire enter to the topological region earlier (i.e.~lower $V_z$) than the middle as $V_z$ is increased. 
We specifically choose the constant potential $\Delta \mu_0 =2$meV and its region length $l_\mu=200$ while the wire length $L=400$. As the Zeeman splitting $V_z$ is increased, the wire regions on the two sides without the additional potential enter the topological phase after the first TQPT point $V_{z1}=4.1$meV. Since the middle of the wire is trivial, four localized MZMs appear separately at the respective wire ends and the potential jump points as illustrated in Fig.~\ref{Step_function_lx200}(a). Effectively, the single wire is now divided into three spatial regimes: two topological regimes with MZMs and one trivial regime with higher chemical potential in the middle. As $V_z$ keeps increasing, the middle region eventually becomes topological at $V_{z2}=\sqrt{0.9^2+(2+4)^2}=6.07$meV at the second TQPT point. As the entire wire is now topological for $V>V_{z2}$, the two internal Majorana modes strongly hybridize away from zero energy and at the same time two MZMs survive on the two wire ends since they are spatially well-separated from each other. The spectrum as a function of $V_z$ (Fig.~\ref{Step_function_lx200}(b,c)) shows the bulk SC gap closing at $V_{z1},\ V_{z2}$, four Majorana modes with small energy splitting between the two gap closings, and two Majorana modes with small energy oscillation for $V_z>V_{z2}$. The energy splitting and oscillation involve the hybridization variations of the Majorana modes as $V_z$ is increased\cite{PhysRevB.86.220506}.

	By tuning the Zeeman splitting $V_z$, the nanowire can host zero, four, and two Majorana modes respectively as three distinct phases. The trivial and topological regions hosting zero and two Majorana modes respectively have been extensively discussed in section \ref{Topo_SC_sec}. Since we have discussed above, our focus now is on the low energy physics of the Josephson effect for the wire hosting four Majorana modes. Before performing the numerical simulation for the lattice model, we construct the low energy Hamiltonian of the four-Majorana model to understand the Josephson junction physics 
\begin{align}
H_{\rm{4M}}=&i(d \cos (\Phi/2) +D)\gamma_l\gamma_r  + i g\gamma_l\lambda_l + ig\gamma_r \lambda_r  \nonumber \\
& + i f \lambda_l \lambda_r +i h\gamma_r \lambda_l + ih \gamma_l \lambda_r , \label{4M effective}
\end{align}
where the Majorana operators ($\gamma_l,\ \gamma_r,\ \lambda_l,\ \lambda_r$) represent two Majoranas located on the two wire ends and two Majoranas located near the potential jump points respectively as illustrated in Fig.~\ref{Step_function_lx200}(a).  The first term in eq.~\ref{4M effective} stems from both the $4\pi$ energy oscillation of the two Majorana end modes and their finite size effect as described in Eq.~\ref{M2}. The second and third terms describe the two similar couplings of the two Majorana modes in the same potential wells (on the left and right sides in Fig.~\ref{Step_function_lx200}(a) respectively) and $f$ is the tunnel coupling strength for the two Majorana modes through the middle of the wire. The last two terms are the coupling of one Majorana end mode on the right/left side and one Majorana mode near the potential jump on the left/right side through the junction. Although $h$ might be $\Phi$-dependent due to the junction tunneling, in our simple model a $\Phi$-independent $h$ is assumed. If the flux-dependence of the coupling strength is known, it is straightforward to include it in the theory.

By analytically solving the eigenvalues of the Hamiltonian, the low energy spectrum of the many-body BCS wavefunction is given by 
\begin{align}
E_{\rm{BCS}1}^\pm= &\pm \sqrt{(d \cos (\Phi/2) +D+f)^2+4g^2}, \nonumber \\
E_{\rm{BCS}2}^\pm =&\pm \sqrt{(d \cos (\Phi/2) +D-f)^2+4h^2} .
\end{align}
These many-body energies lead to the expression of the quasiparticle and quasihole energies 
\begin{align}
E_1^\pm=& \pm E_{\rm{BCS}1}^+ \mp E_{\rm{BCS}2}^+, \nonumber \\
E_2^\pm=& \pm E_{\rm{BCS}1}^+ \pm E_{\rm{BCS}2}^+.
\end{align} 
For our specific model of the numerical simulation, $D\approx 0$ due to the long wire length $L=400$ for all of the following cases. The coupling strength $h$ is neglected, since it is weaken by the long length of the potential well ($(L-l_\mu)/2=100$) in our model and the coupling through the junction (it will be restored for large $l_\mu$ later). The explicit energy expression of quasiparticle and quasihole is written as  
\begin{align}
E_1^\pm=&\pm \sqrt{(d \cos (\Phi/2)+f)^2+4g^2} \mp |d \cos (\Phi/2) -f|,   \nonumber \\
E_2^\pm=&\pm \sqrt{(d \cos (\Phi/2)+f)^2+4g^2} \pm |d \cos (\Phi/2) -f|.  \label{fourM_E}
\end{align} 
The spectrum of Eq.~\ref{fourM_E} shows that non-zero $d,\ f,$ and $g$ destroy the $2\pi$ periodicity ($E_1^\pm(\Phi+2\pi)\neq E_1^\pm(\Phi)$ and $E_2^\pm(\Phi+2\pi)\neq E_2^\pm(\Phi)$) and lead to a $4\pi$ periodicity. This is a key feature that effective $4\pi$ oscillations may arise even when the whole wire encloses multiple MZMs.

	Now returning to the numerical simulation of the $\Phi-E$ relation for the lattice model, since there are four low energy Majorana modes in this region and we include four lowest energy bands in the panels of the $\Phi-E$ relation as shown in Fig.~\ref{Step_function_lx200}(d,e); the $\Phi-E$ relation exhibits a $4\pi$ periodicity in the entire region of the four Majorana modes. Furthermore, the energy spectrum from the lattice model as a function of $\Phi$ in the four-MZM region is consistent with the low energy model of Eq.~(\ref{fourM_E}). We conclude that the transition point between $2\pi$ and $4\pi$ periodicities occurs at the first TQPT point $V_{z1}$ and for $V_z>V_{z1}$, the system manifests only $4\pi$ oscillations in spite of the presence of four MZMs in the wire. 

	The length $l_\mu$ of the additional constant potential region can go to two limits ($l_\mu \rightarrow 0,\ L$). First, as $l_\mu \rightarrow 0$, only the coupling of the two Majorana end modes through the junction and the coupling of the two Majorana modes at the potential jumps dominate so that the effective Hamiltonian is of the simple form 
\begin{align}
H_{\rm{4M}}=&id \cos (\Phi/2) \gamma_l\gamma_r   + i f \lambda_l \lambda_r . \label{f_d}
\end{align}
The coupling strength $f$ grows as $l_\mu$ becomes shorter. When $l_\mu=0$, the two Majorana modes ($\lambda_l,\ \lambda_r$) in the middle of the wire move far away from zero energy and the nanowire hosts only two MZMs ($\gamma_l,\ \gamma_r$) on the ends. For non-zero $l_\mu$, the quasiparticle and quasihole energies as a function of $\Phi$ become similar to Fig.~\ref{Step_function_lx200}(e) ($E_1^\pm =  \pm d \cos (\Phi/2),\ E_2^{\pm}=\pm f$) and the $4\pi$-periodicity appears beyond $V_{z1}$ consistent with expectations (since the system is now simply one homogeneous nanowire).

Second, as $l_\mu \rightarrow L$, $f$ vanishes and the coupling $h$ through the junction between one Majorana end mode and one Majorana mode near one potential jump  on the other side cannot be neglected. These two factors are the key leading to the $2\pi$ periodicity of the BCS wavefunction with fixed fermion parity as shown below. The energy spectrum of quasiparticle and quasihole has the following expression
\begin{align}
E_1^\pm=&\pm \sqrt{d^2 \cos^2 (\Phi/2)+4g^2} \pm \sqrt{d^2 \cos^2 (\Phi/2)+4h^2},  \nonumber \\
E_2^\pm=&\pm \sqrt{d^2 \cos^2 (\Phi/2)+4g^2} \mp \sqrt{d^2 \cos^2 (\Phi/2)+4h^2}.  \label{fourM_E_long}
\end{align} 
The absence of $f$ in the energies implies $E_1^\pm(\Phi+2\pi)=E_1^\pm(\Phi)$ and $E_2^\pm(\Phi+2\pi)=E_2^\pm(\Phi)$. Furthermore, the presence of $g,\ h$ avoids the energy level crossing at $E_1^\pm=E_2^\mp$ as $\Phi=\pi,\ 3\pi$, and without any energy level crossing the Josephson junction does not possess the $4\pi$ periodicity of the BCS wavefunction.

\begin{figure}[t!]
\begin{center}
\includegraphics[clip,width=0.98\columnwidth]{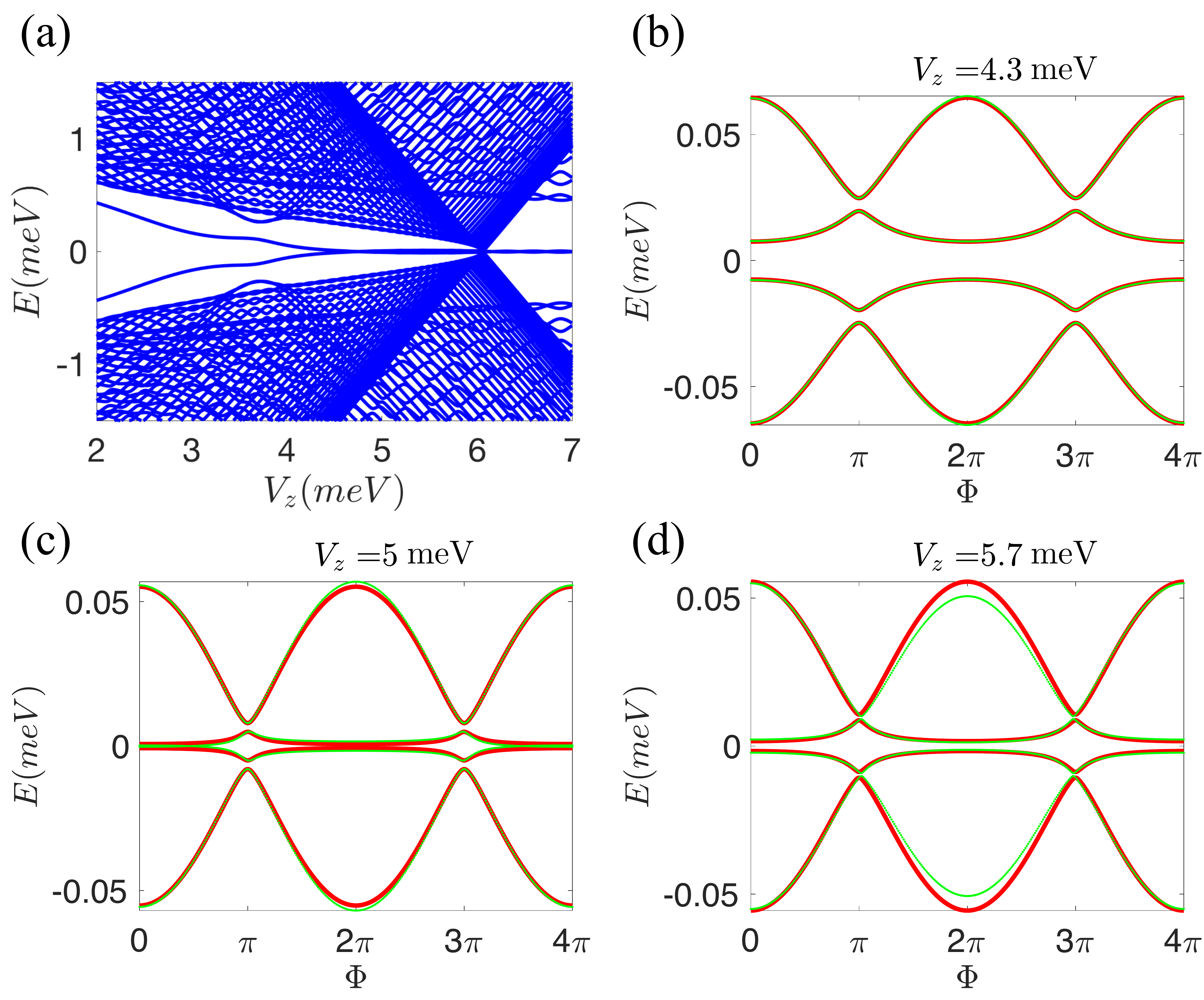}
  \caption{  (a) the energy spectrum for the long length ($l_\mu=350$) of the constant chemical potential plateau ($\Delta \mu_0$) indicates that the bulk gap does not close at $V_z=V_{z1}$ due to the short lengths of the topological regions. (b,c,d) the $\Phi-E$ relations in the lattice model (\ref{lattice Hamiltonian})(green) are in agreement with the effective low energy theory (red, $d,\ g,\ h$ in Eq.~\ref{fourM_E_long} are obtained by fitting). (d) the $\Phi-E$ relation in the lattice model (green) deviates from the effective low energy theory (\ref{fourM_E_long})(red) as $V_z$ is close to $V_{z2}$. That is, the $2\pi$ periodicity gradually changes to the $4\pi$ periodicity as $V_z$ varies from $V_{z1}$ to $V_{z2}$. 
  } 
  \label{lx350}
  \end{center}
\end{figure}

	We consider the specific case $l_\mu=350$ (c.f. $L=400$) for the numerical simulation. The energy spectrum of the wire in the open boundary condition shows that the bulk gap closing does not occur at $V_{1z}$ due to the short length of the topological regions ($(L-l_\mu)/2=25$) as shown in Fig.~\ref{lx350}(a). In the $V_z$ region between $V_{z1}$ and $V_{z2}$, the four energy bands are close to zero energy. We include these four energy bands to calculate the $\Phi-E$ relations of the numerical model as shown in Fig.~\ref{lx350}(b,c,d). As $V_z$ is close to $V_{z1}$, the quasiparticle and quasihole energies (\ref{fourM_E_long}) are identical to the $\Phi-E$ relation from the numerical simulation (Fig.~\ref{lx350}(b)), which exhibits a $2\pi$ periodicity. On the other hand, as $V_z$ is close to $V_{z2}$, the energy spectrum (Fig.~\ref{lx350}(d)) from the numerical simulation deviates away from the low energy approximation (\ref{fourM_E_long}) exhibiting $2\pi$ periodicity. In other words, the portion of the $4\pi$ periodicity in the $\Phi-E$ gradually increases as $V_z$ increases toward $V_{z2}$. 


	In summary, the length of the potential plateau $l_\mu$ and the Zeeman splitting $V_z$ are the main parameters to tune the periodicity of the $\Phi-E$ relation. As $V_z$ is fixed in the region between $V_{z1}$ and $V_{z2}$ and the length $l_\mu$ of the high potential region is increased, the $\Phi-E$ relation gradually changes from a $4\pi$ periodicity to a $2\pi$ periodicity. 
	For large $l_\mu$, as $V_z$ varies from $V_{z1}$ to $V_{z2}$, the periodicity gradually goes from $2\pi$ to $4\pi$, too.  


\begin{figure}[t!]
\begin{center}
\includegraphics[clip,width=0.98\columnwidth]{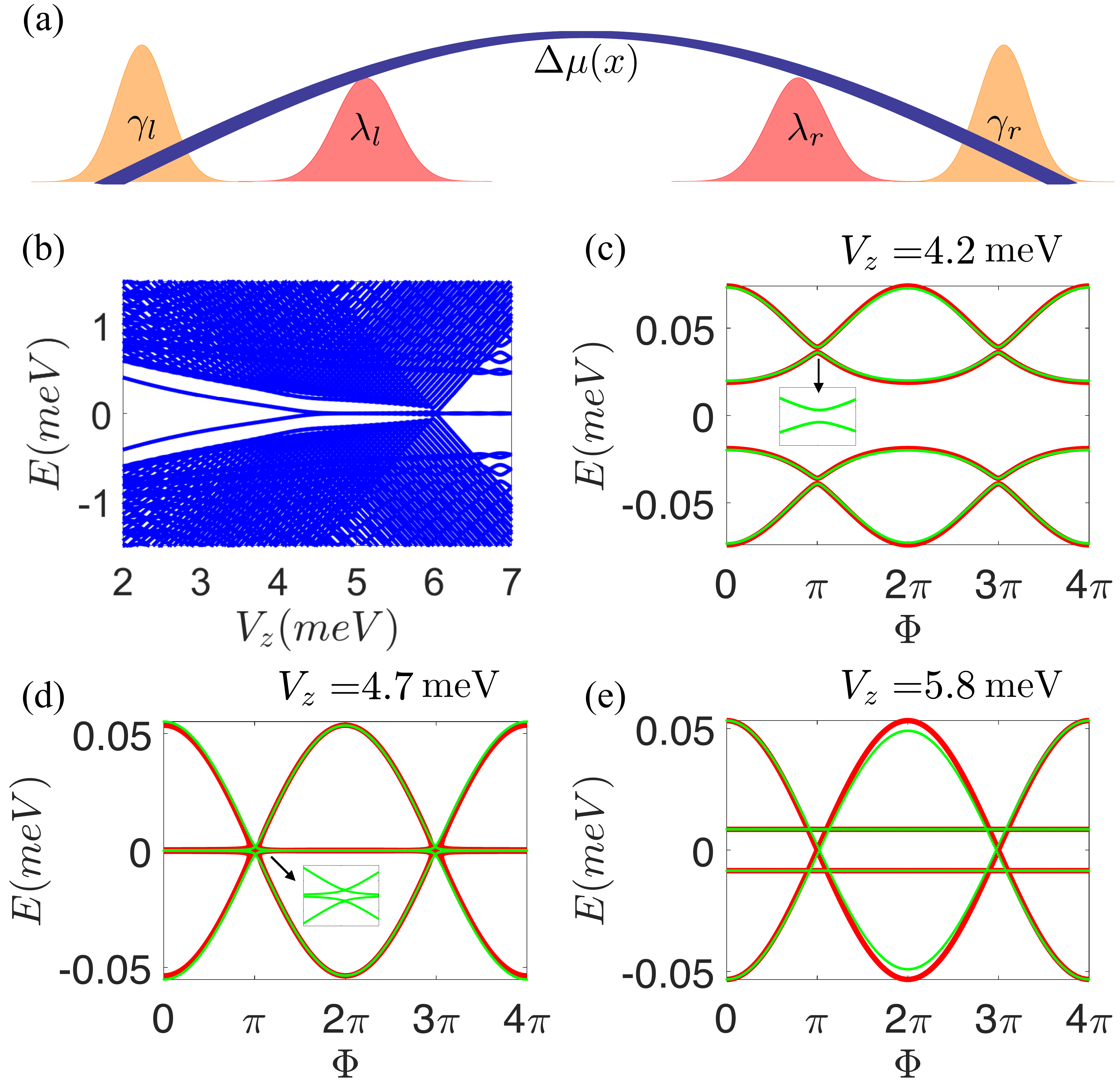}
  \caption{ Panel (a) illustrates a smooth chemical potential distribution $\Delta \mu(x)$ as $\sin x$ in the entire wire. (b) the energy spectrum shows the bulk energy level closing at $V_z=V_{z2}$. Similarly, four Majorana modes appear $(\gamma_l,\lambda_l,\lambda_r,\gamma_r)$ as $V_{z1}<V_z<V_{z2}$ as illustrated in panel (a). (c) the $\Phi-E$ relation at $V_z\sim V_{z1}$ exhibits a $2\pi$ periodicity since the two lowest energy levels are disconnected as shown in the inset. (d) for $V_z$ moves aways from $V_{z1}$, the $\Phi-E$ relation exhibits a $4\pi$ periodicity due to the energy level crossings at $\Phi=\pi,\ 3\pi$ as shown in the inset. The $\Phi-E$ relations (c,d) in the lattice model (green) are consistent with the low energy Majorana model (red, $d,\ h,\ f$ in Eq.~\ref{fourM_E_long} are chosen by fitting). (e) the $\Phi-E$ relation shows a $4\pi$ periodicity as $V_z$ is close to $V_{z2}$; the $\Phi-E$ relation in the lattice model (green) is almost consistent with the two independent energy splittings in the effective theory (red, $d,\ f$ in Eq.~\ref{f_d} are obtained by fitting).} 
  \label{all_sin}
  \end{center}
\end{figure}

\subsection{Sine function potential in all regions}

Next, we consider a smooth trigonometric (``sine") inhomogeneous chemical potential distribution in the entire wire by adding \cite{PhysRevB.97.165302}
\bee
\hat{H}_{\rm{sin}}=-\sum_{1 \leq j \leq L  } V_0 C_j^\dagger \sin(\frac{\pi j}{L})  \tau_z \sigma_0 C_j, 
\ee	
to the lattice Hamiltonian (\ref{lattice Hamiltonian}) as illustrated in Fig.~\ref{all_sin}(a). For the lattice model we choose $V_0=2$meV, which is identical to the constant value of the potential well in the previous case. There are still two distinct TQPT points ($V_z=V_{z1},\ V_{z2}$). The spectrum of the wire (Fig.~\ref{all_sin}(b)) in the open boundary condition shows that two MZMs appear for $V_z>V_{z2}$ and four energy levels close to zero are present between $V_{z1}$ and $V_{z2}$. It is not surprising that the $\Phi-E$ relation of the Josephson effect can still be captured by the four-Majorana effective Hamiltonian (\ref{4M effective}). First, when $V_z$ is near $V_{z1}$ as shown in Fig.~\ref{all_sin}(c), the $\Phi-E$ relation can be faithfully described by Eq.~\ref{fourM_E_long} with non-vanishing $g$ and $h$. Furthermore, the inset of the panel (c) indicates that the two positive energy levels are disconnected. Therefore, the $\Phi-E$ relation exhibits a $2\pi$ periodicity (by assuming the absence of Landau-Zener tunneling). As $V_z$ increases toward $V_{z2}$, the lengths of the topological regions on the two sides become longer to weaken the coupling $h$ between one Majorana end mode and another Majorana near the potential jump on the other side.  With vanishing $h$, Eq.~\ref{fourM_E_long} is in agreement with the $\Phi-E$ relation from the lattice model simulation. 
The $\Phi-E$ relation exhibits a $4\pi$ periodicity since the continuous evolution of the two lowest positive energy levels switch at the energy level crossing (the inset of Fig.~\ref{all_sin}(d)). When $V_{z}$ is very close to $V_{z2}$, the two topological regions are extended and the trivial region the middle of the wire shrinks  significantly. The coupling $f$ between $\lambda_l$ and $\lambda_r$ grows strongly and the coupling $g$ can be neglected due to the long length of the two topological regions. As shown in Fig.~\ref{all_sin}(e), the $\Phi-E$ relation can be simply captured by the two independent energy splittings $d\cos(\Phi/2)$ and $f$ in Eq.~\ref{f_d}.

\begin{figure}[t!]
\begin{center}
\includegraphics[clip,width=0.98\columnwidth]{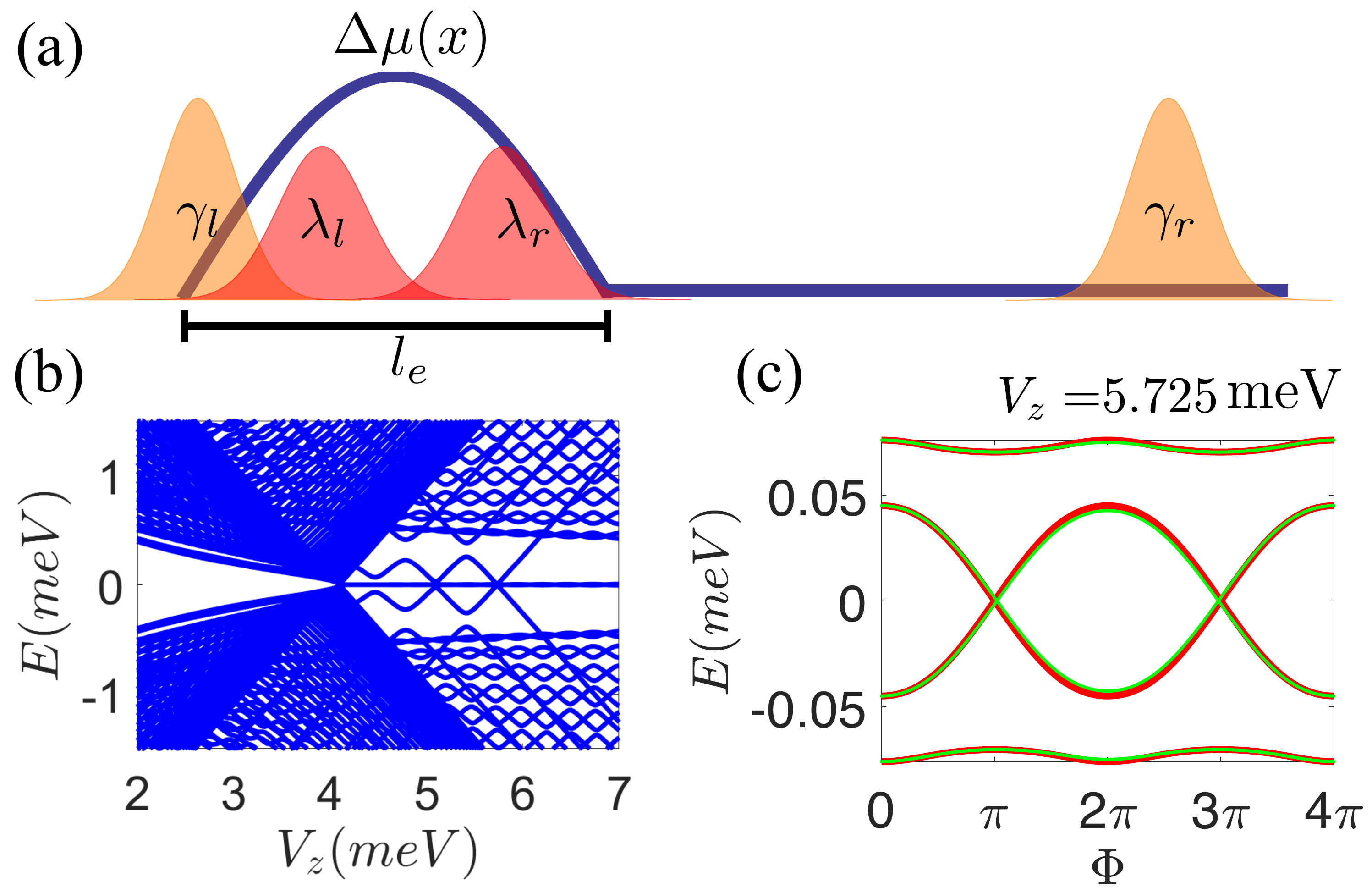}
  \caption{ Panel (a) illustrates an additional sine-shape chemical potential $\Delta \mu(x)$ near one wire end and pictorial locations of the four Majorana modes $(\lambda_l,\gamma_l,\gamma_r,\lambda_r)$. (b) the energy spectrum shows that two MZMs stay at zero energy and two other Majorana modes exhibit the oscillation of the energy splitting as $V_z$ is increased from $V_{z1}$, where $V_{z1}$ is the only TQPT point. (c) the $\Phi-E$ relation of the lattice model (green) is in agreement with the low energy model (red, $d, g, f$ in Eq.~\ref{fourM_E_long_f} are chosen by fitting) and exhibits a $4\pi$ periodicity.  } 
  \label{end_sin}
  \end{center}
\end{figure}

\subsection{Sine function potential near one wire end}

After discussing the smooth potential in the entire wire, we consider a smooth potential variation locally near one end of the wire. As illustrated in Fig.~\ref{end_sin}(a), the additional chemical potential Hamiltonian 
\bee
\hat{H}_{\rm{sin}}=-\sum_{1 \leq j \leq l_e  } V_0 C_j^\dagger \sin(\frac{\pi j}{l_e})  \tau_z \sigma_0 C_j, 
\ee	
is added to the lattice Hamiltonian (\ref{lattice Hamiltonian}). 
In our model, we choose $l_e=50$ and plot the energy spectrum in the open boundary condition in Fig.~\ref{end_sin}(a). Only one TQPT point is located at $V_z=V_{z1}=4.1$meV. Beyond the TQPT point, two low energy modes appear ($\lambda_l,\ \lambda_r$) near the middle of the potential well, and the energy splitting oscillates and moves away from zero energy as $V_z$ is increased, whereas two MZMs $(\gamma_l,\gamma_r)$ are present on the two wire ends separately. We note that for $V_z>V_{z1}$ the energies of the Majorana modes near the middle potential well oscillate and never touch the energy level of the end MZMs. Similarly, the low-energy physics can be captured by this effective four Majorana Hamiltonian
\bee
H_{4M}^{\rm{end}}=id \cos (\Phi/2)\gamma_l\gamma_r  + i g\gamma_l\lambda_l
 + i f \lambda_l \lambda_r , \label{4M end effective}
\ee
where $g$ is the coupling of the Majorana modes on the left side and $f$ is the coupling of the Majorana modes in the middle of the potential well. By solving the algebra of the effective Hamiltonian, the quasiparticle and quasihole energies are given by 
\begin{small}
 \begin{align}
E_1^\pm=&\pm \sqrt{(d \cos (\Phi/2)+f)^2+4g^2} \mp \sqrt{(d \cos (\Phi/2)-f)^2+4g^2},  \nonumber \\
E_2^\pm=&\pm \sqrt{(d \cos (\Phi/2)+f)^2+4g^2} \pm \sqrt{(d \cos (\Phi/2)-f)^2+4g^2}.  \label{fourM_E_long_f}
\end{align} 
\end{small}
These effective low energies are consistent with the $\Phi-E$ relation computed in the lattice model (\ref{lattice Hamiltonian}) as shown in Fig.~\ref{end_sin}(c). Furthermore, the energies exhibit $2\pi$ periodicity with the relations $E_1^\pm(\Phi+2\pi)=E_1^\mp(\Phi)$ and $E_2^\pm(\Phi+2\pi)=E_2^\pm(\Phi)$. The energy level crossing $E_1^\pm=0$ always occurs at $\Phi=\pi,\ 3\pi$. 
Hence, the entire system always exhibits a $4\pi$ periodicity for this local varying potential.

\section{a long trivial superconductor in the middle of the wire} \label{long trivial}

	In the experimental setup\cite{2017arXiv171208459L}, the typical structure of the Josephson device commonly consists of a long trivial (conventional) superconductor in the middle of the wire and the topological superconductors on the two sides of the trivial superconductor; the ends of the two topological superconductors form the junction. Consider the potential distribution exhibiting a plateau in the middle of the wire as shown in Fig.~\ref{Step_function_lx200}(a). As $V_{z1}<V<V_{z2}$, the non-trivial superconducting regions appear on the two sides of the wire and the trivial region is in the middle of the wire. Therefore, the inhomogeneous potential distribution can also faithfully capture the Josephson physics of this long trivial superconductor case. Using the effective four-Majorana model (\ref{4M effective}), we analyze the periodicities of the Josephson effect in this scenario for the different lengths of the topological regions. We note that the two Majorana coupling strengths $D, f$ can be neglected due to the long length of the trivial superconductor in the middle of the wire. (a) when the lengths of the topological superconductors on the two sides are too short, the finite size effect leads to the strong Majorana couplings ($g,\ h$). That is, in the limit of $l_\mu\rightarrow L $, the spectrum of the quasiparticle and quasihole in Eq.~\ref{fourM_E_long} exhibits a $2\pi$ periodicity as $\Phi$ varies. (b) when the lengths of the side topological superconductor regions are long enough, these Majorana hybridizations $g, h$ can be neglected. The low energy Hamiltonian is given by Eq.\ \ref{f_d} without $if\lambda_l\lambda_r$. Therefore, the $E-\Phi$ relation exhibits a $4\pi$ periodicity with additional two isolated Majorana zero modes in the ends of the long trivial superconductors. 
	
	Sec.\ \ref{wire} shows that in the absence of the MZMs, the short length of the wire and the suppression of the superconductivity can lead a $4\pi$ periodicity of the Josephson effect. The long trivial superconductor can exclude some of the trivial $4\pi$-periodicity cases stemming from the short length of the wire. On the other hand, the suppression of the superconducting in this case cannot be ruled out since the $E-\Phi$ still exhibits a $4\pi$ periodicity when the superconductivity in the entire system is suppressed. As the entire system becomes a normal metal, the previous trivial and topological regions does not affect the $4\pi$ periodicity of the $E-\Phi$ relation anymore.

\section{Conclusion} \label{conclusion}

	The presence of isolated MZMs, which possess zero energy, is not the only condition leading to the fractional Josephson effect with $4\pi$ periodicity. When the MZMs are destroyed by the hybridization, a short length superconducting wire and superconducting gap suppression separately may give rise to $4\pi$ periodicity. That is, under some circumstances, the $4\pi$ periodicity may dominate the $E-\Phi$ relation even when the system is not inherently topological. In the topological region, even if the Majorana end modes strongly hybridize due to the finite size effect, the Josephson effect still exhibits a $4\pi$ periodicity. With this strong finite size effect, the energy hybridization never reaches zero as $\Phi$ varies and strictly speaking there is no zero mode in the system Although the experimental setup with the long trivial superconductor can exclude the finite size effect, the other trivial conditions can also lead to the trivial $4\pi$ periodicity. Our work shows that depending on various realistic physical effects (e.g. wire length, gap suppression, Andreev bound states, chemical potential variations), the system may manifest $2\pi$ or $4\pi$ oscillations in the Josephson effect or even a combination of both without clearly establishing any underlying topological physics (or rather emphasizing only some aspects of the topological physics). Therefore, it is difficult to associate the mere observation of approximate $4\pi$ oscillations with the presence of isolated MZMs in the system\cite{2017arXiv171208459L}. Only in the idealized situation of a very long wire with no chemical potential fluctuations or  gap suppression one can identify the presence of $4\pi$ ($2\pi$) oscillations in the Josephson effect as being direct evidence for topological (trivial) superconductivity.
	
	
	For the inhomogeneous potentials, we first discuss the appearance of an ABS in a quantum dot at the wire end. In the trivial region, the presence of the ABS manifests only a $2\pi$ periodicity. Second, we consider the potential distributions creating two separate topological regions and a trivial region in the middle of the nanowire. The four-Majorana model we develop can accurately describe the low-energy physics of the Josephson effect. Unless the length of the topological regions is large, the transition of the periodicity from $2\pi$ to $4\pi$ is a crossover, as $V_z$ is increased from zero. When the topological region is long, there is a clear transition point between $2\pi$ and $4\pi$ periodicities, which is the TQPT point. 
	

	The main goal of this work is to determine the periodicity of the phase-energy ($\Phi-E$) relation in currently studied Majorana nanowires. Our focus has been only on the weak coupling of the junction. On the other hand, the current in the Josephson junction is an essential observable to probe the periodicity of the magnetic flux. For the homogeneous potentials, the two-Majorana model (\ref{theory splitting}) directly shows the Josephson current to be proportional to $\cos (\Phi/2)$ as long as the temperature is much smaller than the second lowest energy level. However, for the inhomogeneous potentials, with multiple low energy levels in the four-Majorana model (\ref{4M effective}), the Josephson current has to be derived based on the Fermi occupation numbers at finite temperature\cite{Kwon2004}. This is a complicated numerical problem, which is better done in the context of specific experimental samples, since all the details of various energy scales (e.g.~temperature, SC gap, tunnel couplings, spin-orbit coupling, ABS energies) become crucial in the calculation of the Josephson current. In the current work, we have focused on the universal physics of the energy-flux relationship in the Josephson effect, and have shown that even this universal physics is strongly affected by many realistic physical mechanisms, which destroy the perceived simplicity of the $2\pi$ versus $4\pi$ Josephson oscillations necessarily reflecting the underlying absence or presence of isolated non-Abelian Majorana zero modes.  Our work establishes that, similar to the zero bias tunneling conductance studies\cite{PhysRevB.96.075161,chiu_blockade}, Josephson effect, by itself, might be incapable of providing decisive information about the topological or trivial nature of the nanowire ground states because of many complicating physical effects invariably occurring in realistic systems. This should be a word of caution for future (or past) experimental claims on this important problem.





\section{acknowledgement} 

The authors thank Yingyi Huang, M. Houzet, C.-X. Liu, and J. D. Sau for discussions. This work is supported by Microsoft and Laboratory for Physical Sciences. C.-K. C. was also supported by the Strategic Priority Research Program of the Chinese Academy of Sciences, Grant No. XDB28000000 during the resubmission stage of this work at the Kavli Institute for Theoretical Sciences.

\bibliographystyle{apsrev4-1}
\bibliography{TOPO}

\end{document}